\documentclass[12pt,preprint]{aastex}
\slugcomment{} 
\shorttitle{Four New Lensed Quasars from the SDSS}
\shortauthors{Oguri}
\begin{document}
\title{Discovery of Four Gravitationally Lensed Quasars from the Sloan
Digital Sky Survey}   
%
\author{
Masamune Oguri,\altaffilmark{1,2}
Naohisa Inada,\altaffilmark{3,4}
Alejandro Clocchiatti,\altaffilmark{5} 
Issha Kayo,\altaffilmark{6}
Min-Su Shin,\altaffilmark{2}
Joseph F. Hennawi,\altaffilmark{7} 
Michael A. Strauss,\altaffilmark{2}
Tomoki Morokuma,\altaffilmark{4,8}
Donald P. Schneider,\altaffilmark{9} and 
Donald G. York\altaffilmark{10} 
}

\altaffiltext{1}{Kavli Institute for Particle Astrophysics and
 Cosmology, Stanford University, 2575 Sand Hill Road, Menlo Park, CA
 94025.} 
\altaffiltext{2}{Princeton University Observatory, Peyton Hall,
 Princeton, NJ 08544.} 
\altaffiltext{3}{Cosmic Radiation Laboratory, RIKEN (The Institute of
 Physical and Chemical Research), 2-1 Hirosawa, Wako, Saitama 351-0198, Japan.} 
\altaffiltext{4}{Institute of Astronomy, Faculty of Science, University
of Tokyo, 2-21-1 Osawa, Mitaka, Tokyo 181-0015, Japan.} 
\altaffiltext{5}{Pontificia Universidad Cat\'{o}lica de Chile,
Departamento de Astronom\'{\i}a y Astrof\'{\i}sica,
Casilla 306, Santiago 22, Chile.}
\altaffiltext{6}{Department of Physics and Astrophysics, Nagoya
University, Chikusa-ku, Nagoya 464-8062, Japan.} 
\altaffiltext{7}{Department of Astronomy, University of California at Berkeley, 601
 Campbell Hall, Berkeley, CA 94720-3411.}
\altaffiltext{8}{Optical and Infrared Astronomy Division, National Astronomical
 Observatory of Japan, Mitaka, Tokyo, 181-8588, Japan.}
\altaffiltext{9}{Department of Astronomy and Astrophysics, The
                  Pennsylvania State University, 525 Davey Laboratory, 
                  University Park, PA 16802.}   
\altaffiltext{10}{Dept. of Astronomy and Astrophysics, and The Enrico
 Fermi Institute, 5640 So. Ellis Avenue, AAC002, The University of
 Chicago, Chicago, IL 60637.}

\begin{abstract}
We present the discovery of four gravitationally lensed quasars
selected from the spectroscopic quasar catalog of the Sloan Digital
Sky Survey. We describe imaging and spectroscopic follow-up
observations that support the lensing interpretation of the following
four quasars: SDSS~J0832+0404 (image separation $\theta=1\farcs98$,
source redshift $z_s=1.115$, lens redshift $z_l=0.659$);
SDSS~J1216+3529 ($\theta=1\farcs49$, $z_s=2.012$); SDSS~J1322+1052
($\theta=2\farcs00$, $z_s=1.716$); and SDSS~J1524+4409
($\theta=1\farcs67$, $z_s=1.210$, $z_l=0.320$). Each system has two
lensed images. We find that the fainter image component of
SDSS~J0832+0404 is significantly redder than the brighter component,
perhaps because of differential reddening by the lensing galaxy. The
lens potential of SDSS~J1216+3529 might be complicated by the presence
of a secondary galaxy near the main lensing galaxy.  
\end{abstract}

\keywords{gravitational lensing --- quasars: general} 

\section{Introduction}\label{sec:intro}

Gravitationally lensed multiple quasars provide unique probes of
astronomical objects, and are also useful to constrain cosmological
parameters. Since gravitational lensing is a pure gravitational
effect, it allows us to determine mass distributions of astrophysical
objects including dark matter.  In addition, the well-known underlying
physics makes it a reliable tool to study cosmology. See
\citet{kochanek06} for a recent review of applications of lensed
quasars. 

The power of lensed quasars as astrophysical and cosmological probes
is enhanced by constructing statistical samples of these objects with
well-defined source populations and well-understood selection
functions. Statistical lens samples have been constructed in the optical
\citep[e.g.,][]{maoz93} and the radio \citep[e.g.,][]{helbig99}
bands. The current largest statistical sample is a radio sample: The
Cosmic Lens All-Sky Survey \citep[CLASS;][]{myers03,browne03} discovered
$~20$ lensed quasars among $~16,000$ radio sources. The sample has been
used to constrain cosmological parameters \citep[e.g.,][]{chae02}.

The purpose of the Sloan Digital Sky Survey Quasar Lens Search
\citep[SQLS;][]{oguri06} is to construct a large statistical lens
sample in optical band. The SQLS searches for 
lensed quasars among spectroscopically confirmed quasar catalogs in the
Sloan Digital Sky Survey \citep[SDSS;][]{york00}. Thus far we have
discovered $>20$ new lensed quasars as well as several previously
known lenses in the SDSS footprint \citep[e.g.,][and references
  therein]{kayo07}, therefore it has already become the largest
statistical sample of strongly lensed quasars. \citet{inada07}
presented the first statistical lens sample of the SQLS from the Data
Release 3 (DR3).

In this paper, we report the discoveries of four additional lensed
quasars from the SQLS. All the systems are double lenses with
small ($1''-2''$) image separations. They are initially selected
from the SDSS data, and the lensing hypothesis is tested against
photometric and spectroscopic follow-up observations at the University
of Hawaii 2.2-meter (UH88) telescope, the United Kingdom Infra-Red
Telescope (UKIRT), the Subaru telescope, the European Southern
Observatory 3.6-meter (ESO3.6m) telescope, and the Astrophysical
Research Consortium 3.5-meter (ARC3.5m) Telescope.  We also perform
simple mass modeling to further test the validity of their lensing
interpretation. Throughout the paper we adopt
the standard cosmology with matter density $\Omega_M=0.27$ and 
cosmological constant $\Omega_\Lambda=0.73$.   The dimensionless
Hubble constant is denoted by $h$. 

\section{Selection of Lens Candidates}\label{sec:sdss}

The aim of the SDSS is to map one quarter of the entire sky by
conducting both a photometric survey \citep{gunn98,lupton99,tucker06} in
five broad-band optical filters \citep{fukugita96} and a spectroscopic
survey with a multi-fiber spectrograph covering 3800{\,\AA} to
9200{\,\AA} at a resolution of $\hbox{R}\sim1800$
\citep{blanton03}. The SDSS uses a dedicated 2.5-m telescope
\citep{gunn06} at the Apache Point 
Observatory in New Mexico, USA.  The data are processed by automated
pipelines \citep{lupton01,lupton07}. The targets for spectroscopy are
selected based on colors and morphology in the imaging survey
\citep{eisenstein01,richards02,strauss02}. The astrometry is accurate
to better than about $0\farcs1$ rms per coordinate \citep{pier03} and
the photometry is calibrated to less than about 0.02 magnitude over 
the entire survey area \citep{hogg01,smith02,ivezic04}. Most of the
data are publicly available
\citep{stoughton02,abazajian03,abazajian04,abazajian05,adelman06,adelman07a,adelman07b}.    

\begin{figure}
\epsscale{.8}
\plotone{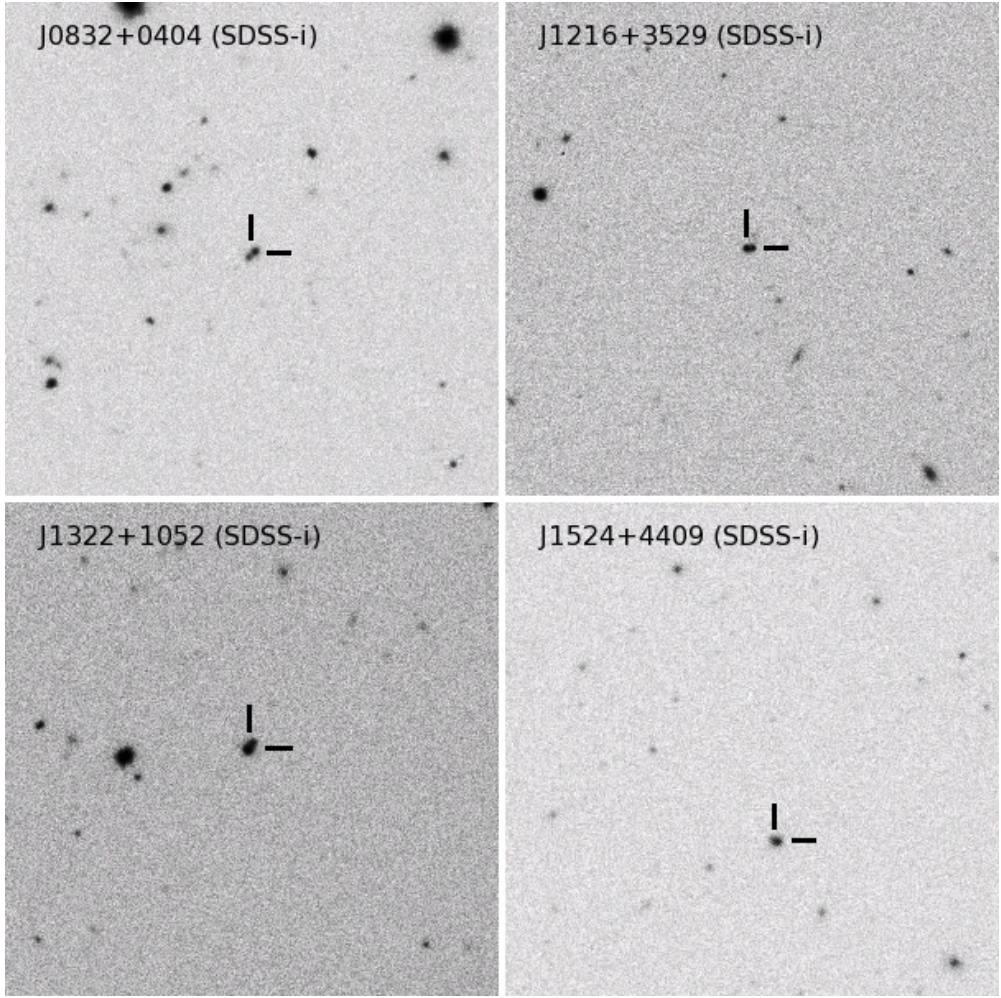}
\caption{Wide-field SDSS $i$-band images of four new lensed quasars,
 SDSS J0832+0404 ({\it upper left}), SDSS  J1216+3529 ({\it upper
 right}), SDSS J1322+1052 ({\it lower left}), and SDSS J1524+4409 ({\it
 lower right}). The size of each image is $2'\times 2'$. North is up and
 East is left.
\label{fig:sdss-chart}}
\end{figure}

Lens systems presented in this paper are selected from the SDSS data
using the algorithm described in \citet{oguri06}. The basic strategy is
to select lens candidates from low-redshift ($z<2.2$) quasars in the SDSS
spectroscopic quasar catalogs
\citep[e.g.,][]{schneider05,schneider07}. We use two 
types of selection methods: Morphological selection, designed to
locate small-separation ($\sim 1''$) lens candidates, and color
selection, to identify lenses with larger image separations.  
The efficiency of our lens selection is $\sim 10\%$ at small
($\theta<2''$) image  separations \citep{oguri06,inada07}.
We summarize the SDSS properties of our new lenses in Table
\ref{tab:sdss}. The SDSS images of these four candidates are shown in
Figure \ref{fig:sdss-chart}. Two of them, SDSS J0832+0404 and SDSS
J1216+3529, are selected by the color selection algorithm, whereas the
other two are identified by the morphological selection algorithm. 

\section{Follow-Up Observations of Individual Objects}\label{sec:follow}

\subsection{SDSS~J0832+0404}\label{sec:0832}
Optical ($V$ and $I$) images of SDSS~J0832+0404 were taken with the
Tektronix 2048$\times$2048 CCD camera (Tek2k; the pixel scale is
$0\farcs2195$ pixel$^{-1}$) at the UH88 telescope on 2006 Nov 15. The
seeing was $\sim 1\farcs1$. Images with a total exposure time for each
filter of 600~sec were taken under photometric conditions. The fluxes
were calibrated using the standard star SA 113 339 \citep{landolt92}.
We also obtained near-infrared ($K$) images of this object with the
UKIRT Fast-Track Imager (UFTI; the pixel scale is $0\farcs091$
pixel$^{-1}$) at UKIRT on 2006 October 13. The seeing was $\sim
0\farcs7$, and the total exposure time was 1080~sec.
The flux is calibrated using the standard star P545-C \citep{persson98}.

Spectral follow-up observations were carried out with the ESO Faint Object
Spectrograph and Camera (EFOSC2) at the ESO3.6m on 2005 December 31. The
total exposure time was 1800~sec. The wavelength coverage was
3700{\,\AA} to 9300{\,\AA} with a spectral dispersion of
$\sim5${\,\AA}~pixel$^{-1}$. The spatial resolution was
$0\farcs31$~pixel$^{-1}$. We used a grating of 236 lines/mm and a
$1\farcs0$ slit. The slit was aligned to observe the two components
simultaneously. The spectral resolution was $R\sim 400$. We extracted
the spectra using standard IRAF\footnote{ 
IRAF is distributed by the National Optical Astronomy Observatories,
    which are operated by the Association of Universities for Research
    in Astronomy, Inc., under cooperative agreement with the National
    Science Foundation.} tasks. We calibrate the spectra
by matching the summed spectral energy distribution (SED) of the two
objects to that of the SDSS spectrum.  Since the angular separation
between the fainter quasar image and the lensing galaxy is small (see
below) we did not separate these two components in studying the
spectra. 

\begin{figure}
\epsscale{.8}
\plotone{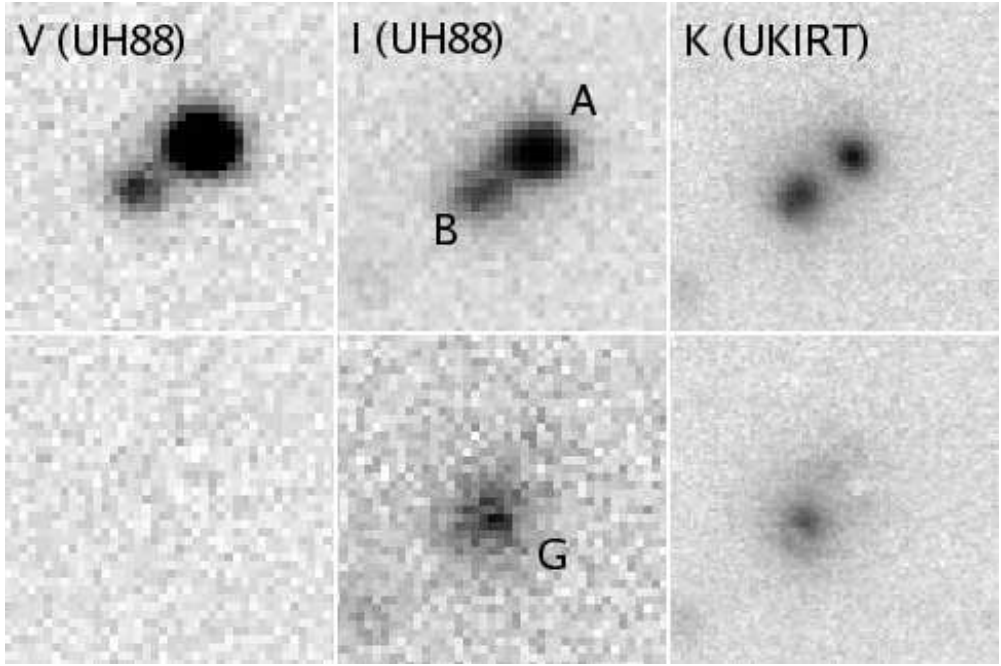}
\caption{Images of SDSS J0832+0404 taken with Tek2k at the UH88
 telescope ($V$ and $I$) and UFTI at UKIRT ($K$). The upper panels
 show original images, and the bottom panels show residuals after
 subtracting two stellar (quasar) components; here the lens galaxy is
 seen more clearly. North is up and East is left, and the size of each
 image is $8\farcs8\times 8\farcs8$. The results are summarized in Table
 \ref{tab:0832}.   
\label{fig:img-sdss0832}}
\end{figure}

\begin{figure}
\epsscale{.8}
\plotone{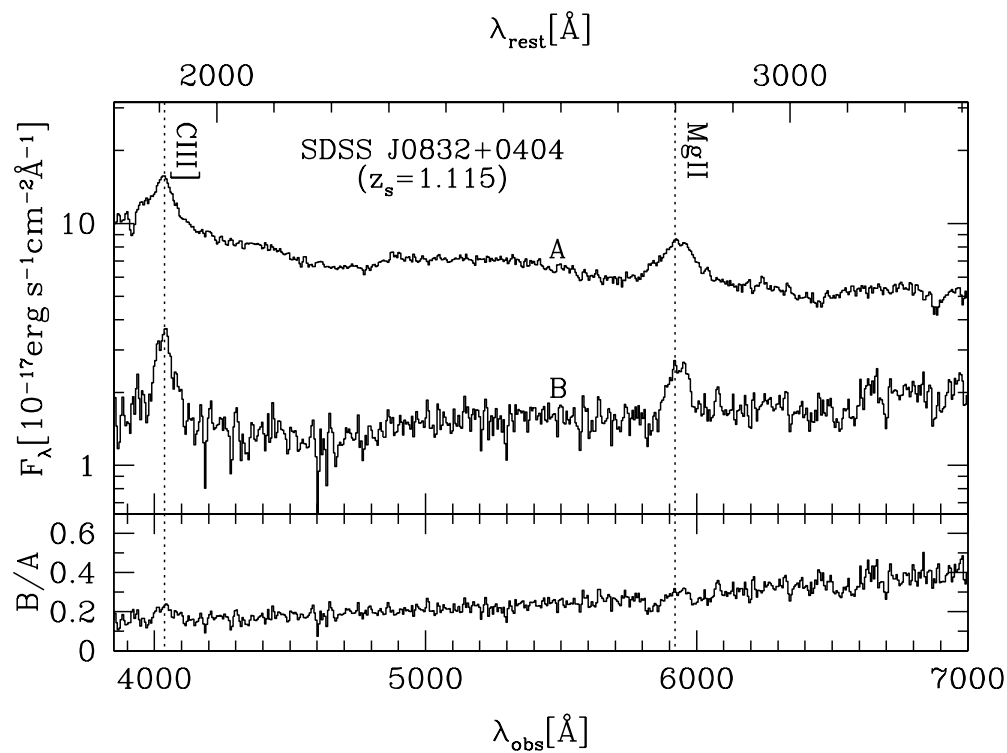}
\caption{Spectra of SDSS J0832+0404 A and B(+G) taken with EFOSC2 at
 the ESO3.6m telescope. The spectral resolution is $R\sim 400$. Dotted
 lines indicate quasar emission lines redshifted to $z_s=1.115$. The
 lower panel plots the ratio of the two spectra.  
\label{fig:spec0832}}
\end{figure}

The results of our follow-up imaging is shown in Figure
\ref{fig:img-sdss0832}. We fit the system using GALFIT
\citep{peng02} with two PSFs plus a galaxy modeled by de Vaucouleurs
profile in $R$ and $I$ band images, and two PSFs in $V$-band image 
because the galaxy is not bright enough. We adopt nearby stars as PSF
templates. Subtracting the two PSFs
shows the presence of a red lensing galaxy more clearly. The fits give
us the relative position and brightness of each component (image A,
image B, and lens galaxy G), which we summarize in Table
\ref{tab:0832}. The image separation between two stellar components is
$1\farcs984\pm0\farcs008$. The spectra shown in Figure
\ref{fig:spec0832} confirm that the two stellar components have broad
emission lines (\ion{C}{3}] and  \ion{Mg}{2}) at the same
wavelengths, supporting the lensing hypothesis of this system. We
note that lensed quasar images (components A and B) 
have different colors; component B ($V-K\sim 3.1$) is considerably redder
than component A ($V-K\sim2.$).  This is probably because of the
differential dust reddening from the lensing galaxy, which is sometimes
seen in lensed quasar systems \citep[e.g.,][]{falco99}.   

\begin{figure}
\epsscale{.8}
\plotone{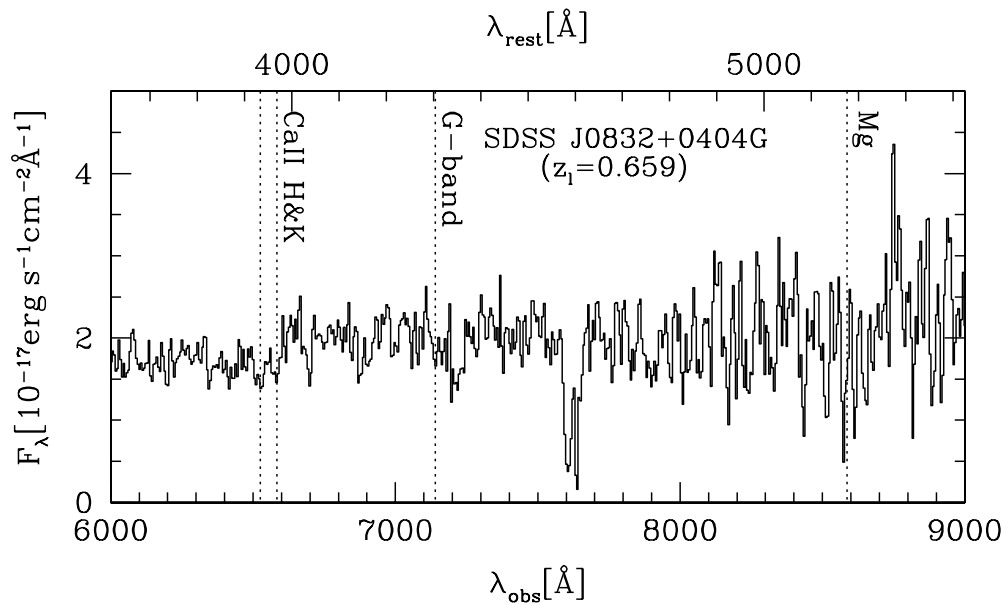}
\caption{The expanded view of the spectrum of SDSS J0832+0404 B
 (Figure \ref{fig:spec0832}) that shows an indication of the lens
 galaxy. The break at $\sim 6600${\,\AA} and two adjacent absorption
 lines suggests that the lens redshift of this system is $z_l=0.659$.
 The strong absorption at $\sim 7600${\,\AA} is telluric. 
\label{fig:spec0832g}}
\end{figure}

We also measure the redshift of the lensing galaxy from our follow-up
spectrum. The spectrum shown in Figure \ref{fig:spec0832g} shows a
break at $\sim6600${\,\AA} (it is weak because of the contamination
from image B) and two adjacent absorption lines that we
interpret as the 4000{\,\AA} break and Ca H \& K lines of an elliptical
galaxy. From the Ca H \& K lines we determine the lens redshift to be
$z=0.659\pm0.001$.  The optical-infrared color of the lens galaxy,
$I-K\sim 3.3$, is broadly consistent with this spectroscopic redshift
\citep[e.g.,][]{nelson01}.   

\subsection{SDSS J1216+3529}\label{sec:1216}
We obtained optical ($V$, $R$, and $I$) images of SDSS J1216+3529 with 
the Tek2k at the UH88 telescope on 2007 April 11. The total exposure
time was 600~sec in $I$ and 300~sec in $V$ and $R$.  The seeing was
$\sim 0\farcs7$.  Photometric calibration was performed using the
standard star PG0918+029 \citep{landolt92}. In addition, near-infrared
($H$) images were taken with the Near-Infrared Camera/Fabry-Perot
Spectrometer (NIC-FPS; the pixel scale is $0\farcs2709$~pixel$^{-1}$)
at the ARC3.5m telescope on 2007 March 7. We obtained a total
exposure of 900 sec with $\sim 0\farcs7$  seeing. Since no standard
star was observed, we perform the photometric calibration using the
Two Micron All Sky Survey (2MASS) data.

The spectra of the two stellar components were obtained with the Wide
Field Grism Spectrograph 2 \citep[WFGS2;][]{uehara04} at the UH88
telescope on 2007 May 13. The grating of 300 lines/mm and the
$0\farcs9$ slit results in a spectral resolution of $R\sim 700$
covering wavelength from 4300{\,\AA} to 10000{\,\AA}. The spectral
dispersion is $\sim 5${\,\AA}~pixel$^{-1}$, and the spatial resolution
is $0\farcs37$~pixel$^{-1}$. The exposure time was 5700~sec. The
observation was conducted under good seeing ($\sim 0\farcs8$), which
makes it rather easy to separate two stellar components. The flux is
calibrated by the standard star Feige 34 \citep{oke90} and the data
were reduced using IRAF. 

\begin{figure}
\epsscale{.98}
\plotone{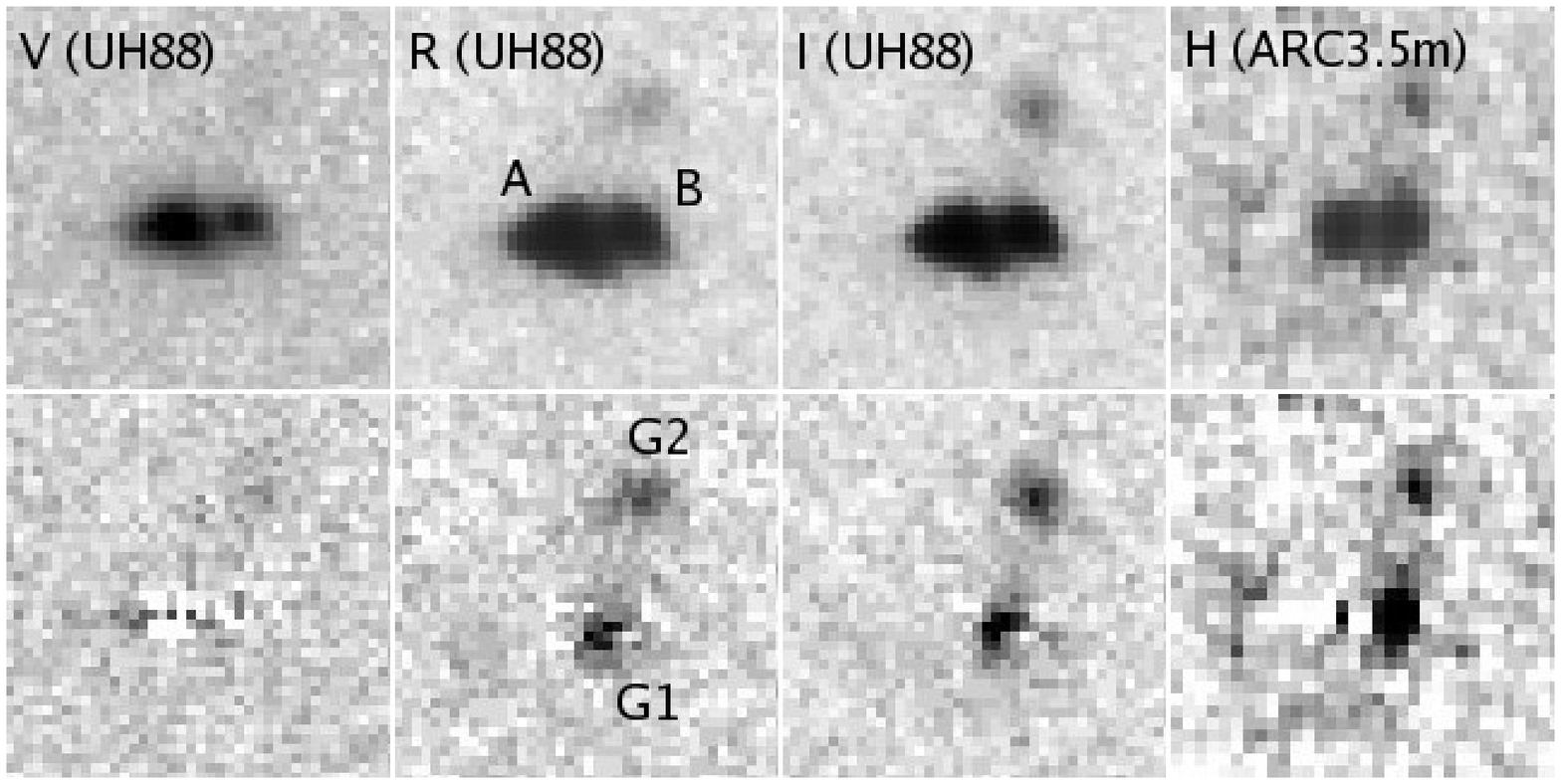}
\caption{Images of SDSS J1216+3529 taken with Tek2k at the
 UH88 telescope ($V$, $R$, and $I$) and NIC-FPS at the ARC3.5m telescope
 ($H$). As in Figure
 \ref{fig:img-sdss0832}, upper and lower panels show original images and
 residuals after subtracting two stellar (quasar) components,
 respectively. North is up
 and East is left, and the size of each image is $8\farcs8\times
 8\farcs8$. The results are summarized in Table \ref{tab:1216}.  
\label{fig:img-sdss1216}}
\end{figure}

We show our results of follow-up imaging in Figure
\ref{fig:img-sdss1216} and Table \ref{tab:0832}. In addition to
the main lensing galaxy (named G1), a red galaxy (denoted 
as G2) is visible north of the system. Thus we fit the system with two
PSFs plus two galaxies (except for $V$-band image in which the
galaxies are too faint) to derive the magnitudes and positions. These
two galaxies may be physically associated, although the colors 
are slightly different such that galaxy G2 is bluer than G1.
The color $R-I\sim 1.0-1.2$ suggests that they are early-type galaxies 
at $z\sim 0.5$ \citep{fukugita95}. The colors of two stellar
components A and B are quite similar. They are separated by an angle
$\theta=1\farcs489\pm0\farcs004$. 

\begin{figure}
\epsscale{.8}
\plotone{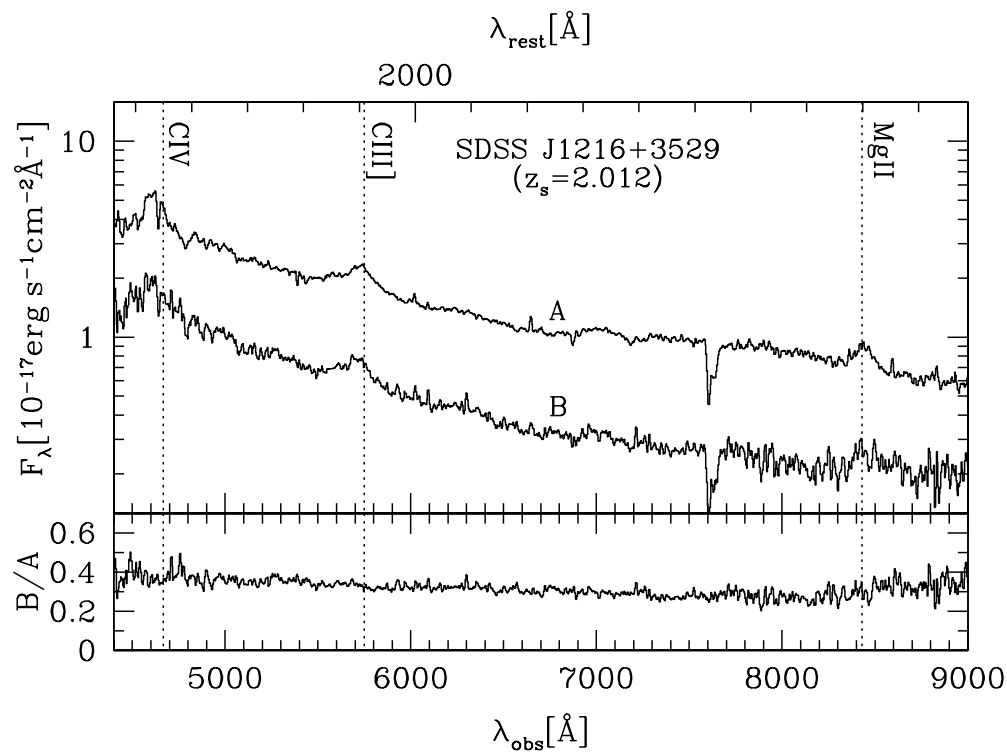}
\caption{Spectra of SDSS J1216+3529 A and B taken with WFGS2 at the UH88
 telescope. The spectral resolution is $R\sim 700$. The spectra are
 smoothed with a 3-pixel boxcar. The location 
 of quasar emission lines redshifted to $z_s=2.012$ is indicated by
 vertical dotted lines. The lower panel plots the ratio of the two
 spectra.  The feature at $\sim 7600${\,\AA} is telluric. 
\label{fig:spec1216}}
\end{figure}

The spectra are shown in Figure \ref{fig:spec1216}. Multiple
quasar emission lines (\ion{C}{4}, \ion{C}{3}], and \ion{Mg}{2}) are
seen in both components A and B, confirming that these are lensed
images of a quasar at $z=2.012$. Moreover the overall shapes of the
spectra are quite similar, as shown in the ratio of the spectra. In 
addition, there are associated absorptions in \ion{C}{4} emission
lines of both components, which further support that these are
gravitationally lensed images. 

\subsection{SDSS J1322+1052}\label{sec:1322}

Optical images of SDSS J1322+1052 were taken with the Tek2k
at the UH88 telescope on 2007 April 11 ($I$; the seeing was
$\sim0\farcs7$) and  2007 May 16 ($V$ and $R$; the seeing was
$\sim0\farcs8$).  The exposure time was 480 sec in $I$ and 300 sec in
$V$ and $R$. The observations were conducted under photometric
conditions, and we used PG0918+029 and PG1633+099 \citep{landolt92} to
derive magnitudes in each image. We took follow-up near-infrared ($H$)
images as well with  NIC-FPS at the ARC3.5m telescope on 2007 April
26. The seeing was $\sim 0\farcs9$ and the total exposure time was
1200~sec. Again, magnitudes are estimated using the 2MASS data.
 
We acquired spectra of this system with the WFGS2 at the UH88
telescope on 2007 May 13. We adopted the same instrument configuration
described in \S \ref{sec:1216}, but with a shorter exposure time of
4500~sec.

\begin{figure}
\epsscale{.8}
\plotone{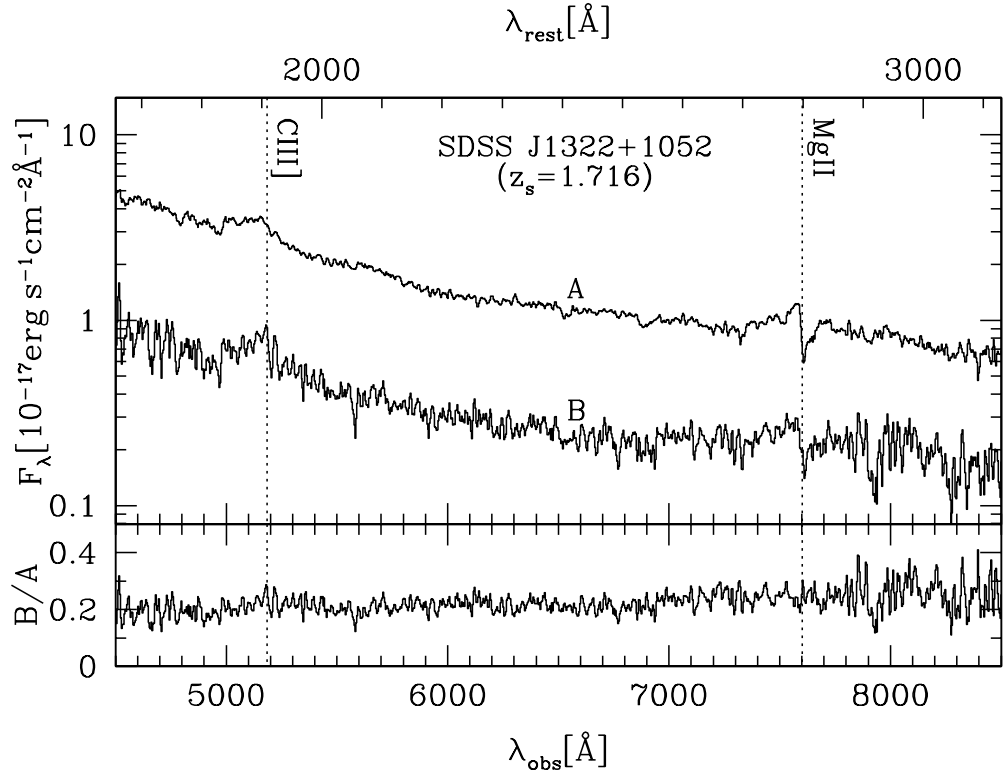}
\caption{Spectra of SDSS J1322+1052 A and B taken with WFGS2 at the UH88
 telescope. The spectral resolution is $R\sim 700$. The spectra are
 smoothed with a 3-pixel boxcar. Vertical 
 lines indicate quasar emission lines redshifted to $z_s=1.716$. Note
 that the \ion{Mg}{2} emission lines are contaminated by telluric
 absorption at $\sim 7600${\,\AA}. The lower panel plots the ratio of
 the two spectra. 
\label{fig:spec1322}}
\end{figure}

Figure \ref{fig:spec1322} presents the follow-up images, and Table
\ref{tab:1322} summarizes the result of fitting. We identify 
a lensing galaxy (component G) in the images in all four bands. The
color, $V-R\sim 1.5$ and $R-I\sim 0.9$, implies a lens redshift of
$z\sim 0.5$. The two stellar components (A and B) have similar colors,
but the magnitude difference is quite large, $\Delta I=1.72$,
corresponding to a flux ratio of $0.205$. The image separation is
$2\farcs001\pm0\farcs007$. The spectra shown in Figure
\ref{fig:spec1322} confirm that components A and B have similar SEDs:
Both components have weak \ion{C}{3}] and \ion{Mg}{2} lines at the
same wavelengths, and the ratio of two spectra is almost constant
over a wide wavelength range. 

\subsection{SDSS~J1524+4409}\label{sec:1524}

We obtained $B$, $R$, and $I$ band images of SDSS~J1524+4409 with
the Orthogonal Parallel Transfer Imaging Camera (OPTIC; the pixel
scale is  $0\farcs1374$~pixel$^{-1}$) at the UH88 telescope on 2006
May 4. The exposure time was 400~sec for each filter and the seeing
was $\sim 1\farcs1$. The standard star PG1633+099 was used to
calibrate the fluxes \citep{landolt92}. 

\begin{figure}
\epsscale{.8}
\plotone{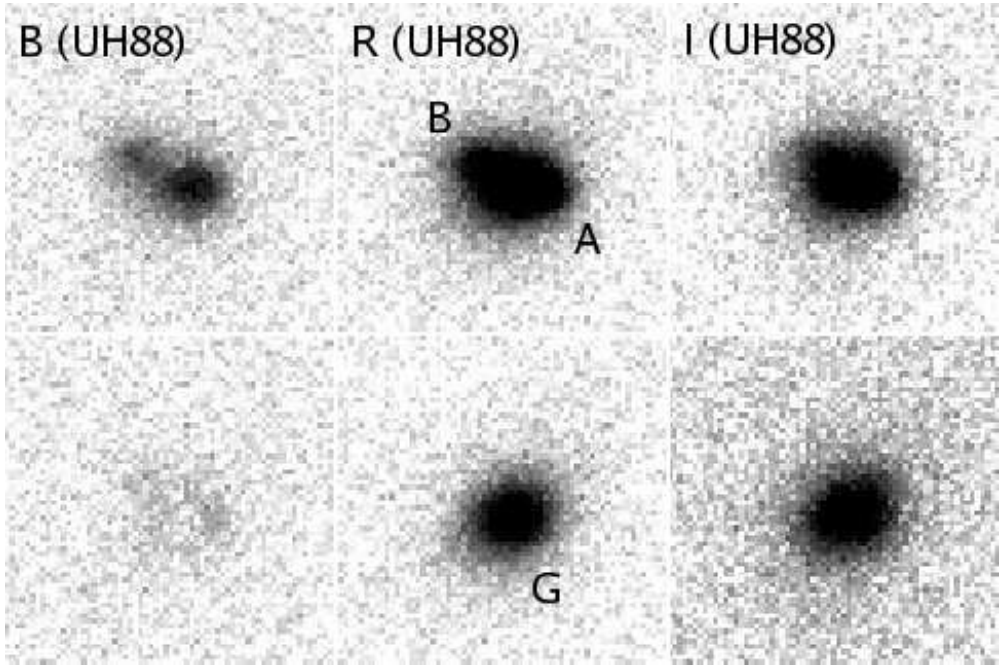}
\caption{Images of SDSS J1524+4409 taken with OPTIC at the UH88
 telescope ($B$, $V$, and $I$). As in Figure \ref{fig:img-sdss0832},
 upper and lower panels 
 show original images and residuals after subtracting two stellar
 (quasar) components, respectively. North is up
 and East is left, and the size of each image is $8\farcs8\times
 8\farcs8$. The result is summarized in
 Table \ref{tab:1524}.  
\label{fig:img-sdss1524}}
\end{figure}

Follow-up images shown in Figure \ref{fig:img-sdss1524} indicate that
this system consists of two stellar components (A and B) and a bright
extended component (G). The colors of components A and B are similar,
thus these are likely to be lensed images. The bright galaxy G
has  red colors and is fitted well by a de Vaucouleurs profile. 
The image separation between A and B is $1\farcs669\pm0\farcs020$. See
Table \ref{tab:1524} for the fitting results.  

The spectrum of this system was taken with the Faint Object Camera and
Spectrograph \citep[FOCAS;][]{kashikawa02} at the Subaru telescope on
2007 January 22. The 300B grism and SY47 filter were used to take the
spectrum in the range from 4700{\,\AA} to 9100{\,\AA} with the
resolution of $R\sim 500$. A long slit with $1''$ width was aligned to
observe the two quasar images simultaneously. The spatial resolution
was $0\farcs2$~pixel$^{-1}$ because of 2$\times$2 on chip binning. The
spectral dispersion after the binning is $\sim 2.6${\,\AA}~pixel$^{-1}$. 
The excellent seeing of $\sim 0\farcs6$ allows us to extract spectra
of all three components (A, B, and G) in a straightforward way.  The
total exposure was 900~sec. The spectra were flux calibrated using the
standard star G191B2B \citep{oke90}. 

\begin{figure}
\epsscale{.8}
\plotone{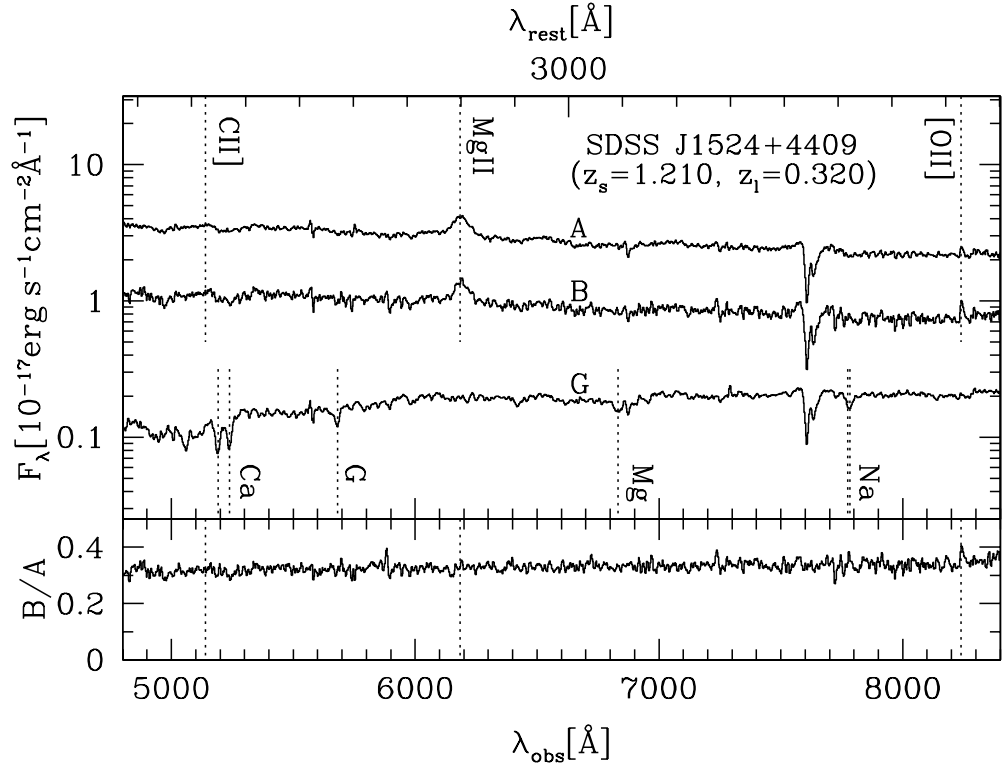}
\caption{Spectra of SDSS J1524+4409 A, B, and G taken with FOCAS at the
 Subaru telescope. The spectral resolution is $R\sim 500$. All the
 spectra are smoothed with a 3-pixel boxcar.  
The spectrum of galaxy G is shifted downward for illustrative purpose. 
Both quasar emission lines redshifted to $z_s=1.210$ and galaxy
 absorption lines redshifted to $z_l=0.320$ are indicated with vertical
 dotted lines.  The feature at $\sim 7600${\,\AA} is telluric.  The
 lower panel plots the ratio of the spectra of components A and B.  
\label{fig:spec1524}}
\end{figure}

The spectra are shown in Figure \ref{fig:spec1524}. As expected,
components A and B have quasar emission lines (\ion{Mg}{2} and
[\ion{O}{2}]) at the same wavelengths and similar overall spectral
shapes, confirming that they are lensed images of a quasar at
$z=1.210$. A number of absorption lines in the spectrum of component G
indicate that the lens is an early-type galaxy at
$z=0.320\pm0.001$. The redshift is in agreement with that expected
from the colors \citep{fukugita95}. 

\section{Mass Modeling}\label{sec:model}

We perform mass modeling of each lens system. Since the lack of
observational constraints for the new lens systems limits detailed
studies of their mass distributions, here we adopt the simplest mass
model, which are commonly used in lens studies, to check whether such
model can fit the new lenses as well. The lens galaxy is assumed
to have a singular isothermal ellipsoid profile parameterized by the
Einstein radius $R_{\rm E}$, ellipticity $e$ and the position angle
(measured East of North) $\theta_e$. We adopt the positions of the quasar
images and lensing galaxies (for SDSS J1216+3529 we consider only
galaxy G1 because of the small number of our observational
constraints) and the fluxes of the quasar images in either $I$ or $K$
band in Tables \ref{tab:0832}-\ref{tab:1524}. The number of degrees of
freedom of this modeling is zero, therefore we should be able to find
models that perfectly fit the observations, as long as the assumed
model is reasonable. Fitting is performed using {\it lensmodel}
package \citep{keeton01}. 

We summarize our results in Table \ref{tab:model}. In all models the
observables are reproduced well, $\chi^2\sim 0$, with reasonable
best-fit values of ellipticity. The existence of feasible mass models
is further support for the hypothesis that these systems are
indeed gravitational lenses.  In Table \ref{tab:model} we show total
magnifications ($\mu_{\rm tot}$) and expected time delays ($\Delta t$)
of the best-fit models. The errors on the time delays from mass model
uncertainties are $\sim 20\%$ \citep{oguri07}.

The mass models also predict velocity dispersions of lens galaxies
that can be converted to luminosities using the Faber-Jackson
relation. We adopt the relation derived by \citet{rusin03} to compute
the expected $I$-band magnitude for each lens system. From the
spectroscopic lens redshifts of SDSS~J0832+0404 and SDSS~J1524+4409,
we calculate the expected magnitudes to be $I=19.4$ and $I=18.9$,
respectively. Given the scatter of the relation, $\sim 0.5$ mag, both
the estimated magnitudes agree with the observed magnitudes of
the lens galaxies. For SDSS J1216+3529 and SDSS J1322+1052, we invert 
the problem and assume the observed lens galaxy magnitudes to
compute the expected lens redshifts from the Faber-Jackson relation.
We find $z_l\sim 0.55$ for both SDSS J1216+3529 and SDSS
J1322+1052. The scatter in the Faber-Jackson relation implies an error
of $\Delta z\sim 0.1$. The estimated redshifts are consistent with the
results inferred from the colors of the lens galaxies. 

Although we were able to fit the systems with a simple singular
isothermal ellipsoid model, it is not clear whether this model really
represents the true density distribution. Complicated lens potentials
could affect lensing statistics \citep[see, e.g.,][for the effect of the
lens galaxy environment]{oguri05}, and thus it is important to explore 
the mass distributions further. The current lack of available number of
observational constrains prevents us from further examinations, hence
additional observations \citep[e.g., the detection of extended lensed
host galaxies; see][]{kochanek01} will be helpful in obtaining more
detailed lens potentials of the individual lens systems. 

\section{Summary}\label{sec:conc}

We have presented the discoveries of four lensed quasar systems. All
were identified as lens candidates on the course of the SQLS, a
systematic survey of lensed quasars from SDSS spectroscopic quasars.
Imaging and spectroscopic observations have shown that the pairs of
stellar images have similar SEDs in each case, and that these  
systems also show lensing galaxies among the stellar components.
The image configurations and image flux ratios were easily reproduced
with simple mass models. Below we summarize the properties of the four
new lensed quasars:

\begin{itemize}
\item SDSS~J0832+0404: A two-image lens with an image separation of
      $\theta=1\farcs984\pm0\farcs008$. The source redshift is
      $z_s=1.115\pm0.003$, and the lens redshift is
      $z_l=0.659\pm0.001$. The fainter component is significantly
      redder than the brighter component.  The $I$-band flux ratio
      $f_I\sim 0.22$ is smaller than the limit for our statistical
      lens sample \citep{oguri06}. 
\item SDSS J1216+3529: A two-image lens with an image separation of
      $\theta=1\farcs489\pm0\farcs004$ and a source redshift of
      $z_s=2.012\pm0.002$. There is a secondary galaxy G2 near the main
      lensing galaxy G1: The rough similarity of their colors suggests
      that they may be physically associated. The observed colors and
      magnitude of the lens galaxy as well as mass modeling suggest
      that the lens redshift is $\sim 0.55$. 
\item SDSS J1322+1052: A two-image lens with an image separation of
      $\theta=2\farcs001\pm0\farcs007$ and a source redshift of
      $z_s=1.716\pm0.002$. Again the $I$-band flux ratio $f_I=0.21$ is
      below the limit for our statistical lens sample. The lens
      redshift is $\sim 0.55$, estimated from colors and magnitude of
      the lens galaxy and mass modeling.
\item SDSS J1524+4409: A two-image lens with an image separation of
      $\theta=1\farcs669\pm0\farcs020$. The source redshift is
      $z_s=1.210\pm0.001$, and the lens redshift is $z_l=0.320\pm0.001$.  
      In $I$-band the lensing galaxy is much brighter than the lensed
      quasar components. 
\end{itemize}

We note that SDSS~J0832+0404 and SDSS J1524+4409 will be included in the
lens lists from the DR3 quasar catalog \citep{inada07}. The DR5 quasar
catalog \citep{schneider07} contains SDSS J1216+3529 and SDSS J1322+1052
as well, but not as lenses. The number of lensed quasars discovered (or
recovered) by the SQLS is now more than 30,\footnote{See
http://www-utap.phys.s.u-tokyo.ac.jp/\~{}sdss/sqls/ for a list of lensed
quasars in the SQLS.} comprising a significant
fraction of all lensed quasars known.

\acknowledgments

This work was supported in part by Department of Energy contract
DE-AC02-76SF00515.
I.~K. acknowledges support from Grant-in-Aid for Scientific Research
on Priority Areas No. 467. 
A.~C. acknowledges the support of CONICYT, Chile, under grant FONDECYT
1051061.
This work is based in part on observations obtained
with the Apache Point Observatory 3.5-meter telescope, which is owned
and operated by the Astrophysical Research Consortium, and on data
collected at Subaru Telescope, which is operated by the National
Astronomical Observatory of Japan.   
Use of the UH 2.2-m telescope and the UKIRT for the observations is
supported by NAOJ. 

Funding for the SDSS and SDSS-II has been provided by the Alfred
P. Sloan Foundation, the Participating Institutions, the National
Science Foundation, the U.S. Department of Energy, the National
Aeronautics and Space Administration, the Japanese Monbukagakusho, the
Max Planck Society, and the Higher Education Funding Council for
England. The SDSS Web Site is http://www.sdss.org/. 

The SDSS is managed by the Astrophysical Research Consortium for the
Participating Institutions. The Participating Institutions are the
American Museum of Natural History, Astrophysical Institute Potsdam,
University of Basel, Cambridge University, Case Western Reserve
University, University of Chicago, Drexel University, Fermilab, the
Institute for Advanced Study, the Japan Participation Group, Johns
Hopkins University, the Joint Institute for Nuclear Astrophysics, the
Kavli Institute for Particle Astrophysics and Cosmology, the Korean
Scientist Group, the Chinese Academy of Sciences (LAMOST), Los Alamos
National Laboratory, the Max-Planck-Institute for Astronomy (MPIA),
the Max-Planck-Institute for Astrophysics (MPA), New Mexico State
University, Ohio State University, University of Pittsburgh,
University of Portsmouth, Princeton University, the United States
Naval Observatory, and the University of Washington.

\clearpage

\begin{deluxetable}{crrcccccc}
\rotate
\tablewidth{0pt}
\tabletypesize{\footnotesize}
\tablecaption{SDSS Properties of Gravitationally Lensed Quasars}
\tablewidth{0pt}
\tablehead{\colhead{Name} & 
 \colhead{RA} &
 \colhead{Dec} &
 \colhead{$u$} & \colhead{$g$} & \colhead{$r$} & \colhead{$i$} &
 \colhead{$z$} & \colhead{Redshift} }
\startdata
SDSS J0832+0404 & 08:32:16.99 & +04:04:05.2 & $19.42\pm0.04$ &
$19.31\pm0.02$ & $18.98\pm0.02$ & $18.95\pm0.02$ & $18.92\pm0.05$ &
$1.115\pm0.003$\\
SDSS J1216+3529 & 12:16:46.05 & +35:29:41.5 & $19.56\pm0.05$ &
$19.43\pm0.03$ & $19.24\pm0.03$ & $19.11\pm0.04$ & $18.85\pm0.04$ &
$2.012\pm0.002$\\
SDSS J1322+1052 & 13:22:36.41 & +10:52:39.4 & $19.82\pm0.04$ &
$19.08\pm0.02$ & $18.75\pm0.01$ & $18.29\pm0.02$ & $18.15\pm0.03$ &
$1.716\pm0.002$\\
SDSS J1524+4409 & 15:24:45.63 & +44:09:49.6 & $19.94\pm0.05$ &
$19.76\pm0.03$ & $19.16\pm0.02$ & $18.82\pm0.02$ & $18.49\pm0.04$ &
$1.210\pm0.001$\\
\enddata
\tablecomments{Magnitudes are Point Spread Function (PSF) magnitudes
 without Galactic extinction correction. The PSF magnitudes roughly
 corresponds to magnitudes of the brighter quasar images for these
 lens systems.}  
\label{tab:sdss}
\end{deluxetable}

\begin{deluxetable}{crrccc}
\tablewidth{0pt}
\tabletypesize{\footnotesize}
\tablecaption{SDSS J0832+0404: Astrometry and Photometry}
\tablewidth{0pt}
\tablehead{\colhead{Name} & 
 \colhead{{$\Delta$}{\rm X}[arcsec]} &
 \colhead{{$\Delta$}{\rm Y}[arcsec]} &
 \colhead{$V$} & \colhead{$I$} & 
 \colhead{$K$} }
\startdata
A & $0.000\pm0.003$  & $0.000\pm0.003$  & $18.87\pm0.01$ & $18.43\pm0.01$ & $16.82\pm0.01$ \\
B & $-1.579\pm0.005$ & $-1.202\pm0.005$ & $20.75\pm0.02$ & $20.08\pm0.08$ & $17.70\pm0.02$ \\
G & $-1.292\pm0.008$ & $-0.875\pm0.008$ & \nodata        & $19.35\pm0.03$ & $16.06\pm0.01$ \\
\enddata
\tablecomments{The positive directions of {$\Delta$}{\rm X} and
 {$\Delta$}{\rm Y} are defined by West and North, respectively. The
 errors are statistical errors only, and do not include systematic
 errors such as model uncertainties, template PSF uncertainties, and
 zero-point errors. The positions are derived in $K$-band image.}  
\label{tab:0832}
\end{deluxetable}

\begin{deluxetable}{crrcccc}
\tablewidth{0pt}
\tabletypesize{\footnotesize}
\tablecaption{SDSS J1216+3529: Astrometry and Photometry}
\tablewidth{0pt}
\tablehead{\colhead{Name} & 
 \colhead{{$\Delta$}{\rm X}[arcsec]} &
 \colhead{{$\Delta$}{\rm Y}[arcsec]} &
 \colhead{$V$} & \colhead{$R$} & 
 \colhead{$I$} & \colhead{$H$} }
\startdata
A & $0.000\pm0.002$  & $0.000\pm0.002$ & $18.97\pm0.01$ & $18.85\pm0.01$ & $18.36\pm0.01$ & $17.30\pm0.01$ \\
B & $1.486\pm0.002$  & $0.097\pm0.002$ & $19.95\pm0.01$ & $19.82\pm0.01$ & $19.33\pm0.01$ & $18.20\pm0.03$ \\
G1& $0.931\pm0.033$  & $-0.075\pm0.033$& \nodata        & $21.58\pm0.06$ & $20.31\pm0.05$ & $18.15\pm0.06$ \\
G2& $1.743\pm0.022$  & $2.891\pm0.022$ & \nodata        & $21.79\pm0.08$ & $20.79\pm0.05$ & $19.23\pm0.13$\\
\enddata
\tablecomments{See Table \ref{tab:0832} for a note.
The positions are derived in $I$-band image.}  
\label{tab:1216}
\end{deluxetable}

\begin{deluxetable}{crrcccc}
\tablewidth{0pt}
\tabletypesize{\footnotesize}
\tablecaption{SDSS J1322+1052: Astrometry and Photometry}
\tablewidth{0pt}
\tablehead{\colhead{Name} & 
 \colhead{{$\Delta$}{\rm X}[arcsec]} &
 \colhead{{$\Delta$}{\rm Y}[arcsec]} &
 \colhead{$V$} & \colhead{$R$} & 
 \colhead{$I$} & \colhead{$H$} }
\startdata
A & $0.000\pm0.002$  & $0.000\pm0.002$ & $18.71\pm0.01$ & $18.41\pm0.01$ & $17.75\pm0.01$ & $16.48\pm0.01$ \\
B & $1.052\pm0.004$  & $1.702\pm0.004$ & $20.50\pm0.02$ & $20.19\pm0.01$ & $19.47\pm0.01$ & $18.04\pm0.02$ \\
G & $0.714\pm0.018$  & $1.169\pm0.018$ & $22.09\pm0.10$ & $20.55\pm0.04$ & $19.67\pm0.03$ & $17.17\pm0.02$\\
\enddata
\tablecomments{See Table \ref{tab:0832} for a note. The positions are
 derived in $I$-band image.}   
\label{tab:1322}
\end{deluxetable}

\begin{deluxetable}{crrccc}
\tablewidth{0pt}
\tabletypesize{\footnotesize}
\tablecaption{SDSS J1524+4409: Astrometry and Photometry}
\tablewidth{0pt}
\tablehead{\colhead{Name} & 
 \colhead{{$\Delta$}{\rm X}[arcsec]} &
 \colhead{{$\Delta$}{\rm Y}[arcsec]} &
 \colhead{$B$} & \colhead{$R$} & 
 \colhead{$I$} }
\startdata
A & $0.000\pm0.008$  & $0.000\pm0.008$ & $20.61\pm0.03$ & $19.98\pm0.02$ & $19.54\pm0.04$ \\
B & $-1.485\pm0.012$ & $0.761\pm0.012$ & $21.36\pm0.05$ & $20.57\pm0.02$ & $20.17\pm0.03$ \\
G & $-0.576\pm0.012$ & $0.142\pm0.012$ & $21.80\pm0.07$ & $19.06\pm0.02$ & $18.33\pm0.01$ \\
\enddata
\tablecomments{See Table \ref{tab:0832} for a note. The positions are
 derived in $I$-band image.}   
\label{tab:1524}
\end{deluxetable}

\begin{deluxetable}{cccccc}
\tablewidth{0pt}
\tabletypesize{\footnotesize}
\tablecaption{Mass Modeling}
\tablewidth{0pt}
\tablehead{\colhead{Name} & 
 \colhead{$R_{\rm Ein}$} &
 \colhead{$e$} & \colhead{$\theta_e$[deg]} & 
 \colhead{$\mu_{\rm tot}$} & \colhead{$\Delta t$[$h^{-1}$day]}}
\startdata
SDSS J0832+0404 & $0\farcs95$ & $0.28$ & $+17$& $3.1$ & $157.8$ \\
SDSS J1216+3529 & $0\farcs77$ & $0.15$ & $-55$& $8.4$ & $15.6$ \\
SDSS J1322+1052 & $1\farcs03$ & $0.18$ & $-33$& $7.3$ & $45.5$ \\
SDSS J1524+4409 & $0\farcs79$ & $0.29$ & $-40$& $7.4$ & $15.3$ \\
\enddata
\tablecomments{For SDSS J1216+3529 and SDSS J1322+1052, we assume the
 lens redshift of $z_l=0.5$ in computing predicted time delays.}  
\label{tab:model}
\end{deluxetable}

\clearpage

\clearpage

\begin{figure}
\epsscale{.98}
\plotone{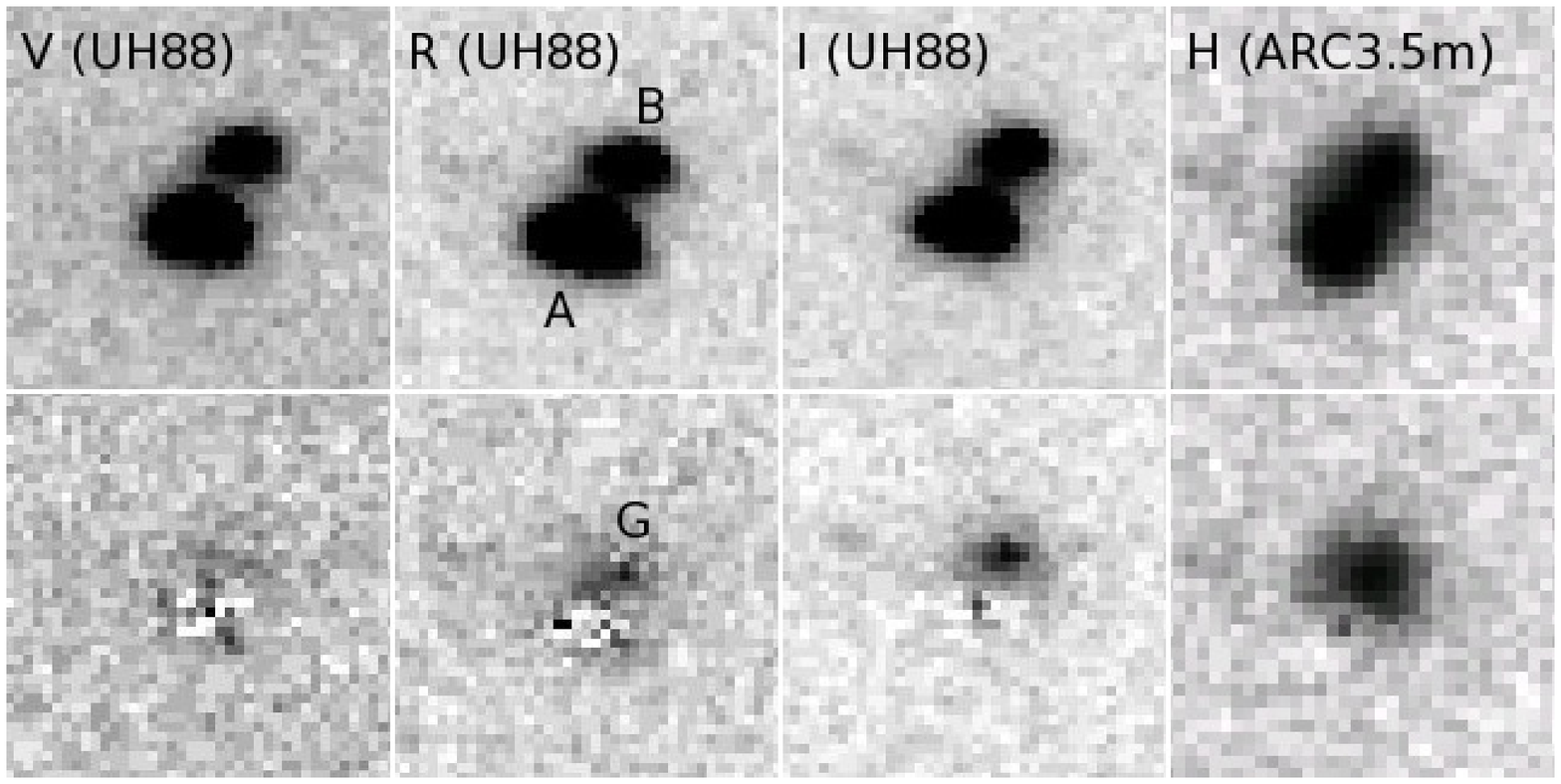}
\caption{Images of SDSS J1322+1052 taken with Tek2k at the 
 UH88 telescope ($V$, $R$, and $I$) and NIC-FPS at the ARC3.5m telescope
 ($H$). As in Figure
 \ref{fig:img-sdss0832}, upper and lower panels show original images and
 residuals after subtracting two stellar (quasar) components,
 respectively. North is up
 and East is left, and the size of each image is $8\farcs8\times
 8\farcs8$. The result is summarized in Table \ref{tab:1322}.   
\label{fig:img-sdss1322}}
\end{figure}

\begin{thebibliography}{}

\bibitem[Abazajian et al.(2003)]{abazajian03}
Abazajian, K., et al. 2003, \aj, 126, 2081

\bibitem[Abazajian et al.(2004)]{abazajian04}
Abazajian, K., et al. 2004, \aj, 128, 502

\bibitem[Abazajian et al.(2005)]{abazajian05}
Abazajian, K., et al. 2005, \aj, 129, 1755

\bibitem[Adelman-McCarthy(2006)]{adelman06}
Adelman-McCarthy, J.~K., et al. 2006, 
\apjs, 162, 38
 
\bibitem[Adelman-McCarthy(2007a)]{adelman07a}
Adelman-McCarthy, J.~K., et al. 2007a, 
\apjs, 172, 634

\bibitem[Adelman-McCarthy(2007b)]{adelman07b}
Adelman-McCarthy, J.~K., et al. 2007b, 
\apjs, submitted (arXiv:0707.3413)
 
\bibitem[Blanton et al.(2003)]{blanton03} 
Blanton, M. R., Lin, H., Lupton, R. H., Maley, F. M., Young, N., 
Zehavi, I., \& Loveday, J. 2003, \aj, 125, 2276

\bibitem[Browne et al.(2003)]{browne03} 
Browne, I. W. A., et al. 2003, \mnras, 341, 13

\bibitem[Chae et al.(2002)]{chae02}
Chae, K.-H., et al.\ 2002, Phys. Rev. Lett., 89, 151301 

\bibitem[Eisenstein et al.(2001)]{eisenstein01} 
Eisenstein, D. J., et al. 2001, \aj, 122, 2267

\bibitem[Falco et al.(1999)]{falco99}
Falco, E.~E., et al.\ 1999, \apj, 523, 617 

\bibitem[Fukugita et al.(1996)]{fukugita96}
Fukugita, M., Ichikawa, T., Gunn, J. E., Doi, M., 
Shimasaku, K., \& Schneider, D. P. 1996, \aj, 111, 1748

\bibitem[Fukugita et al.(1995)]{fukugita95} 
Fukugita, M., Shimasaku, K., \& Ichikawa, T. 1995, PASP, 107, 945

\bibitem[Gunn et al.(1998)]{gunn98} 
Gunn, J. E., et al. 1998, \aj, 116, 3040

\bibitem[Gunn et al.(2006)]{gunn06} 
Gunn, J. E., et al. 2006, \aj, 131, 2332

\bibitem[Helbig et al.(1999)]{helbig99}
Helbig, P., Marlow, D., Quast, R., Wilkinson, P.~N., 
Browne, I.~W.~A., \& Koopmans, L.~V.~E.\ 1999, \aaps, 136, 297 

\bibitem[Hogg et al.(2001)]{hogg01}
Hogg, D. W., Finkbeiner, D. P., Schlegel, D. J., 
\& Gunn, J. E.\ 2001, \aj, 122, 2129

\bibitem[Inada et al.(2007)]{inada07}
Inada, N. et al. 2007, \aj, in press (arXiv:0708.0828)

\bibitem[Ivezi\'{c} et al.(2004)]{ivezic04}
Ivezi\'{c}, \v{Z}., et al. 2004, AN, 325, 583

\bibitem[Kashikawa et al.(2002)]{kashikawa02}
Kashikawa, N., et al.\ 2002, \pasj, 54, 819 

\bibitem[Kayo et al.(2007)]{kayo07}
Kayo, I. et al. 2007, \aj, 134, 1515

\bibitem[Keeton(2001)]{keeton01}
Keeton, C.~R.\ 2001b, preprint (astro-ph/0102340)

\bibitem[Kochanek et al.(2001)]{kochanek01}
Kochanek, C.~S., Keeton, C.~R., \& McLeod, B.~A.\ 
2001, \apj, 547, 50 

\bibitem[Kochanek(2006)]{kochanek06}
Kochanek, C. S., Schneider, P., Wambsganss, J., 2006, 
Part 2 of Gravitational Lensing: Strong, Weak \& Micro, 
Proceedings of the 33rd Saas-Fee Advanced Course, 
G. Meylan, P. Jetzer \& P. North, eds. (Springer-Verlag: Berlin), 91

\bibitem[Landolt(1992)]{landolt92}
Landolt, A. U. 1992, \aj, 104, 340 

\bibitem[Lupton et al.(1999)]{lupton99}
Lupton, R. H., Gunn, J. E., \& Szalay, A. S. 1999, \aj, 118, 1406

\bibitem[Lupton et al.(2001)]{lupton01}
Lupton, R., Gunn, J. E., Ivezi\'c, Z., Knapp, G. R.,
Kent, S., \& Yasuda, N. 2001, in ASP Conf. Ser. 238,
Astronomical Data Analysis Software and Systems X,
ed. F. R. Harnden, Jr., F. A. Primini, and H. E. Payne
(San Francisco: Astr. Soc. Pac.), p. 269 (astro-ph/0101420)

\bibitem[Lupton(2007)]{lupton07}
Lupton, R. 2007, \aj, submitted 

\bibitem[Maoz et al.(1993)]{maoz93}
Maoz, D., et al.\ 1993, \apj, 409, 28 

\bibitem[Myers et al.(2003)]{myers03} 
Myers, S. T., et al. 2003, \mnras, 341, 1

\bibitem[Nelson et al.(2001)]{nelson01}
Nelson, A.~E., Gonzalez, A.~H., Zaritsky, D., \& 
Dalcanton, J.~J.\ 2001, \apj, 563, 629 

\bibitem[Oguri et al.(2005)]{oguri05}
Oguri, M., Keeton, C.~R., \& Dalal, N.\ 2005, \mnras, 364, 1451

\bibitem[Oguri et al.(2006)]{oguri06}
Oguri, M., et al. 2006, \aj, 132, 999

\bibitem[Oguri(2007)]{oguri07}
Oguri, M.\ 2007, \apj, 660, 1 

\bibitem[Oke(1990)]{oke90}
Oke, J.~B.\ 1990, \aj, 99, 1621 

\bibitem[Peng et al.(2002)]{peng02}
Peng, C. Y., Ho, L. C., Impey, C. D., \& 
Rix, H.-W. 2002, \aj, 124, 266  

\bibitem[Persson et al.(1998)]{persson98}
Persson, S.~E., Murphy, D.~C., Krzeminski, W., Roth, M., \& 
Rieke, M.~J.\ 1998, \aj, 116, 2475

\bibitem[Pier et al.(2003)]{pier03} 
Pier, J. R., Munn, J. A., Hindsley, R. B., Hennessy, G. S., 
Kent, S. M., Lupton, R. H., \& Ivezi\'{c}, \'{Z}. 2003, 
\aj, 125, 1559

\bibitem[Richards et al.(2002)]{richards02} 
Richards, G. T., et al. 2002, \aj, 123, 2945

\bibitem[Rusin et al.(2003)]{rusin03}
Rusin, D., et al.\ 2003, \apj, 587, 143 

\bibitem[Schneider et al.(2005)]{schneider05}
Schneider, D.~P., et al.\ 2005, \aj, 130, 367 

\bibitem[Schneider et al.(2007)]{schneider07}
Schneider, D.~P., et al.\ 2007, \aj, 134, 102

\bibitem[Smith et al.(2002)]{smith02} 
Smith, J. A., et al. 2002, \aj, 123, 2121

\bibitem[Stoughton et al.(2002)]{stoughton02}
Stoughton, C., et al. 2002, \aj, 123, 485

\bibitem[Strauss et al.(2002)]{strauss02} 
Strauss, M. A., et al. 2002, \aj, 124, 1810

\bibitem[Tucker et al.(2006)]{tucker06} 
Tucker, D. L., et al. 2006, AN, 327, 821

\bibitem[Uehara et al.(2004)]{uehara04}
Uehara, M., et al.\ 2004, \procspie, 5492, 661 

\bibitem[York et al.(2000)]{york00}
York, D. G., et al. 2000, \aj, 120, 1579

\end{thebibliography}
\end{document}